\documentclass[
 pre,
 aps,
 a4paper,
 english,
 showkeys,
 reprint,
 twocolumn,
 superscriptaddress
]{revtex4}
\usepackage[T1]{fontenc}
\usepackage[utf8]{inputenc}
\usepackage{amsmath}
\usepackage{amssymb}
\usepackage{graphicx}
\usepackage{skak}
\usepackage{multirow}
\usepackage{epsdice}
\usepackage{refstyle}
\usepackage{tabularx}
\usepackage{array}
\usepackage{multirow}
\usepackage[shortlabels]{enumitem}
\usepackage[
citecolor=blue,
colorlinks,
linkcolor=blue,
urlcolor=blue,
]{hyperref}
\usepackage[dvipsnames]{xcolor}
\usepackage{dcolumn}
\usepackage{bm}
\newcommand{\indsize}{\scriptsize}
\newcommand{\colind}[2]{\displaystyle\smash{\mathop{#1}^{\raisebox{.5\normalbaselineskip}{\indsize #2}}}}
\newcommand{\rowind}[1]{\mbox{\indsize #1}}
\begin{document}

\preprint{APS/123-QED}

\title{Selection-recombination-mutation dynamics: Gradient, limit cycle, and closed invariant curve}

	\author{Suman Chakraborty}
	\email{suman@iitk.ac.in}
	\affiliation{
		Department of Physics,
		Indian Institute of Technology Kanpur,
		Uttar Pradesh 208016, India
	}
	\author{Sagar Chakraborty}
	\email{sagarc@iitk.ac.in}
	\affiliation{
		Department of Physics,
		Indian Institute of Technology Kanpur,
		Uttar Pradesh 208016, India
	}


\date{\today}

\begin{abstract}
In this paper, the replicator dynamics of the two-locus two-allele system under weak mutation and weak selection is investigated in a	
generation-wise non-overlapping unstructured population of individuals mating at random. Our main finding is that the dynamics is gradient-like when the point mutations at the two loci are independent. This is in stark contrast to the case of one-locus multi-allele where the existence gradient behaviour is contingent on a specific relationship between the mutation rates. When the mutations are not independent in the two-locus two-allele system, there is the possibility of non-convergent outcomes, like asymptotically stable oscillations, through either the Hopf bifurcation or the Neimark--Sacker bifurcation depending on the strength of the weak selection. The results can be straightforwardly extended for multi-locus two-allele systems.
\end{abstract}

\maketitle

\section{introduction}
\label{sec:intro}
How a living being appears and behaves is inherently determined by the variety of genes located on the loci of chromosomes. The entire genotype to phenotype map is very vast and complex, and is further complicated by the environmental feedback effects. While the evolutionary dynamics---e.g., replicator dynamics~\cite{Taylor1978mathbio, Schuster1983jtb, Cressman2014, Mukhopadhyay2021chaos}---of simplified one-locus many-allele systems exhibits quite rich dynamical behaviour,  the phenomenon of recombination during meiosis presents one with richer extension of the evolutionary dynamics in a population of sexually mating organisms. In the latter, the simplest non-trivial mathematical setup is that of a two-locus two-allele system. While two different loci can determine two different phenotypic traits (e.g., seed colour and seed colour in the revolutionary dihybrid cross experiments of Mendel~\cite{o1993genetic, Mendel1866versuche}), there are plethora of examples where two loci control the same trait and their interaction has considerable effect on the phenotype. Some such examples that quickly comes to mind are comb-types on the head of chicken~\cite{Bateson1909springer}, flower colour in peas~\cite{Dooner1991arg}, wheat kernel color~\cite{carver2009book}, blood group in humans~\cite{dean2005book}, age-related hearing loss resistance in the Japanese wild-derived inbred MSM/Ms mice~\cite{Yasuda2022Biomedicines}, and eye colour of humans~\cite{2010WhiteJHG}.  

Haldane~{\cite{haldane1931mathematical}} and Wright~{\cite{wright1935Genetics}} were among the firsts to mathematically explore the evolutionary outcome when selection acts on more than one locus assumed statistically independent, i.e., system is in linkage equilibrium. Later Lewontin and Kojima~{\cite{lewontin1960evolution}} worked on general 2L2A model with both selection and recombination included, and showed that strong epistasis together with linkage disequilibrium can lead to significantly different outcomes. Subsequently, many extensions~{\cite{bodmer1967genetics, karlin1970TPB, feldman1979genetics, hastings1985genetics, ewens1968TAG}} of the model were studied---starting from prediction and calculation of all possible fixed points  and determination of stability condition for some special cases in simplified fitness model to the occurrence of cyclic motion. For continuous time model, Akin~{\cite{akin1982JMB}} and for discrete time model, Hastings~{\cite{hastings1981PNAS}}, Hofbauer and Iooss {\cite{hofbauer1984Mathematik}} showed that periodic orbits can be a possible outcome when system is in linkage disequilibrium; when the 2L2A model is in linkage equilibrium such complex behaviour can not occur and the system become gradient-like as shown by Nagylaki~{\cite{nagylaki1989Genetics}}. Nagylaki also showed that for multi-locus system under weak selection---selection  strength sufficiently smaller compared to recombination---linkage disequilibrium decays close to zero within a few generations and the dynamics of the entire multilocus system is governed by the dynamical outcomes of time-continuous multi-locus system~\cite{nagylaki1993Genetics,nagylaki1999JMB,nagylaki2013book}. Very recently~{\cite{pontz2018JMB}}, Pontz and coworkers studied the 2L2A model in detail under weak selection limit and classified all possible equilibrium structure and phase portraits for different payoff structures.      

Mutation is the most important and indispensable ingredient of the evolutionary processes. It is important to note that the relationship between mutations at different loci is often complex and can be influenced by a variety of factors. It is possible for a mutation at one locus to affect the mutation rate at another locus on the same chromosome \cite{Nowak2004pnas,Li2013nb} directly or indirectly.  For example, a mutation in a gene that codes for a repair enzyme involved in maintaining the integrity of DNA could alter the mutation rate of other genes by affecting the efficiency of DNA repair~\cite{Steele1998bjs}. Similarly, a mutation in a gene that regulates the transcription or translation of other genes can affect the rate of mutations in those genes. Mutation may not be always completely random: It can depend on environmental changes \cite{Galhardo2007crbmb} and fitness~\cite{Baer2008plos}, and organisms can have evolved mechanisms which can influence the timing or genomic location of mutation~\cite{Rando2007cell}. In view of the above, in this paper, we distinguish between the cases of mutations that occur randomly (independent of what mutations occur at the other loci) from the cases where mutations can be treated to have occurred independently and randomly at a locus---we call the latter independent mutations.

The effects of mutation was studied in one-locus two-allele model for both overlapping and non-overlapping generations, and the results are really interesting: Mutation can stop the extinction of cooperators by producing cyclic or chaotic outcomes and even can produce coexistence regions where both the cooperators and defectors can survive simultaneously~\cite{Mobilia2010jtb,toupo2015PRE,mittal2020pre,mukhopadhyaysuman2021JPhysComplexity}. Evolutionary dynamics were studied for the case of the 2L2A with mutation as well~\cite{park2011JMB,qu2020FrontierGenetics,bengtsson1983TPB,Jain2011jnmp,Brger1989genetics,christiansen1977ajhg,Karlin1971tpb}. Effect of weak selection in 2L2A model, in the absence of mutation, was studied in detail by Pontz et al.~{\cite{pontz2018JMB}}. If the selection is sufficiently weak then the linkage disequilibrium disappears after a few generations and dynamics become quite straightforward to find its equilibrium structure for the general fitness matrix. The system behaves like a gradient system and if all the fixed point are hyperbolic then for any initial condition dynamics ultimately approaches fixed points: No periodic or chaotic outcome can arise. On the other hand, the effect of both weak mutation and weak selection was studied for one-locus many-allele model by Hofbauer and Sigmund~\cite{hofbauer1998book}. They have shown that the corresponding system becomes gradient system when the mutation matrix satisfies certain condition: When mutation probability dependents on the target gene only~\cite{Hofbauer1985jmb}, the system behaves like a gradient system, and no periodic or chaotic orbit is possible. In general, however, the system need not be a gradient system, and hence periodic and chaotic orbits can be possible outcome. 

In such contexts, the study of the 2L2A model in presence of weak selection and weak mutation is still, to the best of our knowledge, remains to be reported. Specifically, we are interested in whether the 2L2A model is always gradient-like or  there is a specific condition on mutation for that to happen. Does the answer depend on whether the mutations at the two locus are independent? Furthermore, if and when the epistasis induced dynamics in 2L2A system be altered due to presence of mutation is also a question of interest because epistasis or interaction between different loci has important effects on dynamics, like occurrence of cyclic motion~\cite{hastings1981PNAS} and sustaining polymorphism~\cite{pontz2018JMB}.

Without further ado, we introduce the model in the next section (Sec.~\ref{section_2}) and then we present the gradient 2L2A system in Sec.~\ref{section_3}. Subsequently, we discuss the non-gradient dynamics in Sec.~\ref{section_3} before concluding in Sec.~\ref{section_4}.

\section{The 2L2A Model}\label{section_2} 
We consider a generation-wise non-overlapping, unstructured population of randomly mating diploid individuals. The population is assumed to evolve under the action of viability selection that acts on two diallelic, recombining loci. We, thus, have the standard two-locus two-allele (2L2A) model with viability selection~\cite{pontz2018JMB,Karlin1975tpb,Karlin1975numericaltpb,Sacker2003jdea,Hallgrmsdttir2008genetics,Lewontin1988tpb,Franklin1977tpb,bodmer1967genetics}. Let $A_1$ and $A_2$ be the alleles at locus $A$, and $B_1$ and $B_2$ be the alleles at locus $B$. Let $G_1$, $G_2$, $G_3$, and $G_4$ represent the four possible gametes, viz., $A_1B_1$, $A_1B_2$, $A_2B_1$, and $A_2B_2$ respectively. Let the frequency of the $G_i$ gametes be $x_i$ (obviously, $\sum_{i=1}^4x_i=1$). {\color{black}In these notations}, the  frequencies of allele \(A_1\) and \(B_1\) are respectively \(x_1+x_2\) and \(x_1+x_3\) that we henceforth denote as $p$ and $q$ respectively. Defining \(D\equiv x_1x_4-x_2x_3\), the measure of linkage disequilibrium, following identities follow:
\begin{subequations}
\label{eqn:frequencis of gametes}
\begin{eqnarray}
&&x_1=pq+D,\\
&&x_2=p(1-q)-D,\\
&&x_3=(1-p)q-D,\\
&&x_4=(1-p)(1-q)+D.
\end{eqnarray}
\end{subequations}

Furthermore, let the fitness of the $G_iG_j$ genotype be $w_{ij}$. For simplicity, we assume that the fitness is independent of the fact which gamete is from mother and which one is from father, i.e., $w_{ij}=w_{ji}$. We also  assume that $w_{14}=w_{23}$, the last assumption means that the corresponding fitnesses are the same regardless of whether $A_1$ and $B_1$ are on the same chromosome or opposite ones. Consequently,  the following matrix suffices for fully representing the fitnesses of genotypes:
\[
\begin{array}{@{}c@{}}\label{matrix_3}
\rowind{$A_1A_1$} \\ \rowind{$A_1A_2$} \\ \rowind{$A_2A_2$}  
\end{array}
\mathop{\left[
	\begin{array}{ *{5}{c} }
	\colind{w_{11}}{$B_1B_1$}  &  \colind{w_{12}}{$B_1B_2$}  &  \colind{w_{22}}{$B_2B_2$}  \\
	w_{13} &  w_{14}  &  w_{24} \\
	w_{33}  & w_{34} &  w_{44} 
	\end{array}
	\right].}^{
	\begin{array}{@{}c@{}}
	\\ \mathstrut
	\end{array}
}
\]

We include two further phenomena in the model: recombination and mutation. While we let $r$ denote the probability of recombination, the mutation from one gamete to another gamete is specified by the row stochastic matrix ${\sf Q}$ whose ($i,j$)th element $Q_{ij}$ is the probability that an offspring with gamete $G_j$ is born to a parent with gamete $G_i$. With this multiplicative~\cite{toupo2015PRE,hofbauer1998book} mutation that takes place during DNA replication process, the evolution of gamete frequencies is given by the following equation~\cite{bengtsson1983TPB,PritchettEwing1981genetics,christiansen1977ajhg,Karlin1971tpb}:  
\begin{equation}
    \label{eqn:x_i}
     x'_i=\frac{\sum_{j}\left[x_jw_j-\theta_j r Dw_{14}\right]Q_{ji}}{\bar w},
\end{equation}
where \(w_i\equiv\sum_jw_{ij}x_j\), \(\bar w\equiv\sum_j w_jx_j\), \(\sum_jQ_{ij}=1\), and $\theta_1=-\theta_2=-\theta_3=\theta_4=1$. Here, prime is the tag for immediately succeeding generation. We suppose that the mutation happens at the gametic stage~\cite{bengtsson1983TPB, Bergero2021biologicalreviews}.

As elaborated in the introduction to this paper, since our goal is to observe effect of both weak selection and weak mutation on the dynamics of the 2L2A selection-recombination model, we introduce a small parameter $s$ such that the fitnesses and mutations can be recast as
\begin{subequations}
	\label{eqn:selection}
	\begin{eqnarray}
	&&w_{ij}=1+sm_{ij},\\&& Q_{ij}=s\epsilon_{ij}\ \ \  {\rm for}\, i\neq j.
	\end{eqnarray}
\end{subequations}
Thus, {\color{black} the limit of small $s$} renders the dynamics of Eq.~(\ref{eqn:x_i}) to be that under weak selection and weak mutation.

It is well-known~{\cite{hofbauer1998book}} that recombination drives 2L2A model to Wright manifold in the absence of selection and mutation: The Wright manifold is a linkage equilibrium manifold ($\Lambda_0$), where $D=0$, and a population in linkage equilibrium always remains in equilibrium, i.e. it is an invariant manifold. In the presence of weak selection $(s\ll r)$ and weak mutation, on using Eq.~(\ref{eqn:selection}) in Eq.~(\ref{eqn:x_i}), we obtain
\begin{equation}
    \label{eqn:x_i with weak selection}
    x'_i=x_i-\theta_i rD+\mathcal{O}(s).
\end{equation}
 Interestingly, we find that the effect of mutation at this order is completely absent. Hence, the known results~\cite{nagylaki2013springer,nagylaki1999springer} for the case of the models without mutation (but with selection) holds in our case as well: Close to $\Lambda_0$, there exists a smooth invariant manifold, $\Lambda_s$, that is globally attracting for Eq.~(\ref{eqn:x_i}); for any initial values the linkage disequilibrium $D(t)$ becomes $\mathcal{O}(s)$ asymptotically in time.

Thus, in the weak selection and weak mutation limit, the linkage disequilibrium $D \to 0$, and consequently, the dynamical equation is further simplified because of the fact that it can now be specified using only two variables $p$ and $q$, for $t \ge t_1$. Specifically, in the set of Eqs.~(\ref{eqn:frequencis of gametes}), $D$ is replaced by $\mathcal{O}(s)$ terms.
  Rescaling time $t (=0,1,2...)$ tagging generations as $\tau=st$, it is easy to see that as $s \to 0$, Eq.~(\ref{eqn:x_i}) approaches the following differential equations:
 \begin{subequations}\label{2}
 	\begin{eqnarray}
 		\dot{p}&=&{q}{(1-p)} \nu_1+{(1-q)}{(1-p)} \nu_2-{q}{p}\nu_3-{(1-q)}{p}\nu_4\nonumber\\ && +p(1-p)\frac{1}{2}\frac{\partial \bar{m}}{\partial p},\\
 		\dot{q}&=&{p}{(1-q)} \nu_5+{(1-p)}{(1-q)} \nu_6-{p}{q}\nu_7-{(1-p)}{q}\nu_8  \nonumber \\&&+q(1-q)\frac{1}{2}\frac{\partial \bar{m}}{\partial q}.
 	\end{eqnarray}
 \end{subequations}
where, $\nu_1\equiv{(\epsilon_{31}+\epsilon_{32})},~ \nu_2\equiv{(\epsilon_{41}+\epsilon_{42})},~ \nu_3\equiv{(\epsilon_{13}+\epsilon_{14})},~ \nu_4\equiv{(\epsilon_{23}+\epsilon_{24})},~
 \nu_5\equiv{(\epsilon_{21}+\epsilon_{23})},~ \nu_6\equiv{(\epsilon_{41}+\epsilon_{43})},~ \nu_7\equiv{(\epsilon_{12}+\epsilon_{14})},~ \nu_8\equiv{(\epsilon_{32}+\epsilon_{34})}$, and
 \begin{equation}
 \bar{m}\equiv m_1pq+m_2p(1-q)+m_3(1-p)q+m_4 (1-p)(1-q)
 \end{equation}
 is the average fitness of the population wherein the marginal fitness of the gamete $i$,
 \begin{equation}
 m_i\equiv m_{i1}pq+m_{i2}p(1-q)+m_{i3}(1-p)q+m_{i4} (1-p)(1-q).
 \end{equation}
{\color{black}Since, at $p=0$ and $p=1$,} $\dot{p}\ge 0$ and $\dot{p} \le 0$ respectively; and at $q=0$ and $q=1$,  $\dot{q}\ge 0$ and $\dot{q} \le 0$ respectively, Eq.~(\ref{2}) is forward invariant in the $p$-$q$ phase space---a closed unit square. It may also be noted that the dynamical equation remain invariant under addition (but not multiplication) of a constant matrix with the matrix ${\sf M}$ whose elements are $m_{ij}$'s. It may be remarked that had the mutation been taken as additive rather than multiplicative~\cite{Mobilia2010jtb,toupo2015PRE}, the form of the final dynamical equation, in the limit of weak selection and weak mutation, remains same as Eqs.~(\ref{2}).
 
\section{2L2A Gradient System}\label{section_3}
If a dynamical system is gradient-like, then the non-convergent solutions like periodic or chaotic solutions can be ruled out the system. So it is very useful to find the condition on the mutation rates for which the 2L2A model becomes gradient-like.

It is of notational convenience to represent Eqs.~(\ref{2}) in the form: 
\begin{eqnarray}
\dot{z}_i=\tilde{f}_i({\boldsymbol{z}})\equiv z_i(1-z_i)f_i({\boldsymbol{z}}),\,\, i=1,2,3,4;\label{2'}
\end{eqnarray}
here, $z_1\equiv p$, $z_2\equiv q$, $z_3\equiv (1-p)$ and $z_4\equiv(1-q)$ correspond to the four coordinates in the allele frequency space. The explicit expression of $f_i$'s is obvious when compared with Eqs.~(\ref{2}). The phase space $\mathbb{P}$, hence, is ${\Sigma}^2\times\Sigma^2$ embedded in $\mathbb{R}^4$; here, $\Sigma^2$ is one dimensional simplex. We need to consider the interior of $\mathbb{P}$, ${\rm int}\mathbb{P}$ (say); at any point ${\boldsymbol{z}}\in {\rm int}\mathbb{P}$, let the tangent space be denoted by $\mathbb{T}_{\boldsymbol{z}}\mathbb{P}$. 

Now for ${\boldsymbol{z}}\in {\rm int}\mathbb{P}$ and $\boldsymbol{\eta}\in \mathbb{T}_{\boldsymbol{z}}\mathbb{P}$, the inner product 
\begin{eqnarray}
\langle\dot{\boldsymbol{z}}, \boldsymbol{\eta} \rangle_{\boldsymbol{z}}=\langle\boldsymbol{\tilde{f}}(\boldsymbol{z}), \boldsymbol{\eta} \rangle_{\boldsymbol{z}}=g_{ij}(\boldsymbol{z})\tilde{f}_i(\boldsymbol{z}) \eta_j
\end{eqnarray}
where (and henceforth) sum over repeated indices---Einstein's summation convention---has been imposed. $g_{ij}$ is the metric in ${\rm int}\mathbb{P}$. 
Considering the metric $g_{ij}={\delta_{ij}}/[z_i(1-z_i)]$---reminiscent of the Shahshahani gradient~\cite{hofbauer1998book,shahshahani1979ams, ebeling1985lotka}---and recalling Eq.~(\ref{2'}), it is obvious that
 \begin{eqnarray}\label{eq:dotz1}
 \langle\dot{\boldsymbol{z}}, \boldsymbol{\eta} \rangle_{\boldsymbol{z}}=\frac{\partial U}{\partial z_i}\eta_i,
 \end{eqnarray}
 if there exists a continuous and differentiable scalar function $U({\boldsymbol{z}})$ such that $f_i(\boldsymbol{z})={\partial U({\boldsymbol{z}})}/{\partial z_i}$. Actually, since $\sum_{i=1}^4\eta_i=0$ because $\boldsymbol{\eta}\in \mathbb{T}_{\boldsymbol{z}}\mathbb{P}$,  even if a more general condition, viz., $f_i(\boldsymbol{z})={\partial U({\boldsymbol{z}})}/{\partial z_i}+\phi({\boldsymbol{z}})$ ($\phi$ is a scalar function) is satisfied, Eq.~(\ref{eq:dotz1}) holds good (see Appendix~\ref{sec:mlma}).
 
In summary, what we have found is that if $\dot{z}_i={f}_i({\boldsymbol{z}})$ is gradient system, then so is $\dot{z}_i=\tilde{f}_i({\boldsymbol{z}})$. Consequently, for Eqs.~(\ref{2}) to be gradient system, one requires
\begin{equation}\label{3}
\frac{\partial f_1}{\partial q}=\frac{\partial f_2}{\partial p}.
\end{equation}
This condition, on putting the expression of $f_i$'s, ultimately takes the form
\begin{equation}\label{4}
\frac{1}{p}(\nu_1-\nu_2)+\frac{1}{(1-p)}(\nu_3-\nu_4)=\frac{1}{q}(\nu_5-\nu_6)+\frac{1}{(1-q)}(\nu_7-\nu_8).
\end{equation}
As $p$ and $q$ are independent variables, the system is gradient system for all possible allowed values of $p$ and $q$ if
	\begin{eqnarray}\label{5}
	&& \nu_1=\nu_2,\,\nu_3=\nu_4,\, \nu_5=\nu_6,\,{\rm and}\,\nu_7=\nu_8.
	\end{eqnarray}
Evidently, the condition on mutation for 2L2A model  to be gradient-like is completely different from the one-locus four-allele model~\cite{hofbauer1998book}.

In the light of condition (\ref{5}), Eqs.~(\ref{2}) take the form
\begin{subequations}
\label{6}
\begin{eqnarray}
\dot{p}=p(1-p)\frac{1}{2}\frac{\partial \bar{m}}{\partial p} +\nu_1(1-p)-\nu_3 p,\\
\dot{q}=q(1-q)\frac{1}{2}\frac{\partial \bar{m}}{\partial q} +\nu_5(1-q)-\nu_7 q.
\end{eqnarray}
\end{subequations}
It is interesting to note that with  $V(p,q)\equiv\exp (\bar{m})p^{2\nu_1}(1-p)^{2\nu_3}q^{2\nu_5}(1-q)^{2\nu_7}$ , Eq.~(\ref{6}) can be recast as: 
\begin{subequations}\label{eq:7}
	\begin{eqnarray}
	&& \dot{p}=\frac{p(1-p)}{2V}\frac{\partial V}{\partial p},\\&&\dot{q}=\frac{q(1-q)}{2V}\frac{\partial V}{\partial q};
	\end{eqnarray}
\end{subequations}
the interior fixed points $(p^*,q^*)$ correspond to  ${\partial V}/{\partial p}={\partial V}/{\partial q}=0$. In fact, it can be verified (see Appendix~\ref{sec:llf}) that $V(p,q)$ is local Lyapunov function for the gradient system (\ref{6}) because
\begin{equation}\label{9}
\begin{aligned}
\dot{V}=\frac{2V}{p(1-p)}\dot{p}^2+\frac{2V}{q(1-q)}\dot{q}^2,
\end{aligned}
\end{equation}
implying $\dot{V}= 0$ only at any $(p^*,q^*)\in(0,1)^2$ and  $\dot{V} > 0$ always for all possible $(p,q)\in(0,1)^2$ other than the fixed points.

The existence of the Lyapunov function leads to the conclusion for Eq.~(\ref{eqn:x_i}): For suffciently small $s$ and if all the equilibria of Eqs.~(\ref{eq:7}) are hyperbolic, every solution of Eq.~(\ref{eqn:x_i}) converges to a fixed point. This is essentially as extension of a theorem due to Nagyalaki~\cite{nagylaki1999JMB} and can be succinctly understood as follows. Substituting $w_{ij}=1+sm_{ij}$ and $Q_{ij}=s\epsilon_{ij}$ for all $i \ne j$ in equation (\ref{eqn:x_i}), and assuming linkage equilibrium ($D=0$) and condition (\ref{5}), we arrive at
\begin{subequations}\label{eq:11}
\begin{eqnarray}
\Delta p=\frac{s}{\bar{w}} \left[ \frac{p(1-p)}{2}\frac{\partial \bar{m}}{\partial p}+\nu_1(1-p)-\nu_3 p\right], \\
\Delta q=\frac{s}{\bar{w}} \left[ \frac{q(1-q)}{2}\frac{\partial \bar{m}}{\partial q}+\nu_5(1-q)-\nu_7 q\right],
\end{eqnarray}
\end{subequations}
which reduce to Eqs.~(\ref{eq:7}) as $s\rightarrow0$. Here, $\bar{w}=1+s\bar{m}$. It is straightforward to note that the fixed points of Eqs.~(\ref{eq:7}) and Eqs.~(\ref{eq:11}), and their stability properties for small enough $s$ are same. This means that if the fixed points of Eqs.~(\ref{eq:7}) are hyperbolic, then due to the system being a gradient system, the initial conditions will be attracted to some stable fixed points; and so is going to be the fate of the initial conditions under the map given by Eq.~(\ref{eqn:x_i}) whose corresponding fixed points are in $\Lambda_s$ manifold.

The search for which kind of replicator-mutator 2L2A system can be a gradient system interestingly stops at the most simple case of random point mutation, happening independent of what is happening elsewhere (e.g., at another locus): We call such mutations independent. It can be expressed in our setup as follows: Let the probability of mutation from $A_i$ allele to $A_j$ allele be given by $\mu_{ij}^A$ and the probability of mutation from $B_i$ allele to $B_j$ allele is given by $\mu_{ij}^B$. Naturally $\mu_{ii}^A$ and $\mu_{ii}^B$ correspond to the probabilities of no mutation from $A_i$ and $B_i$ alleles respectively. So the corresponding mutation matrix ${\sf Q}$ representing the probability of mutation from one gamete to another gamete, when the mutation at the two locus are independent, can be represented by the following matrix:
\[
\begin{array}{@{}c@{}}\label{matrix_1}
\rowind{$A_1B_1$} \\ \rowind{$A_1B_2$} \\ \rowind{$A_2B_1$} \\ \rowind{$A_2B_2$} 
\end{array}
\mathop{\left[
	\begin{array}{ *{5}{c} }
	\colind{\mu_{11}^A \mu_{11}^B}{$A_1B_1$}  &  \colind{\mu_{11}^A \mu_{12}^B}{$A_1B_2$}  &  \colind{\mu_{12}^A \mu_{11}^B}{$A_2B_1$}  & \colind{\mu_{12}^A \mu_{12}^B}{$A_2B_2$}  \\
	\mu_{11}^A \mu_{21}^B &  \mu_{11}^A \mu_{22}^B  &  \mu_{12}^A \mu_{21}^B  & \mu_{12}^A \mu_{22}^B \\
	\mu_{21}^A \mu_{11}^B  & \mu_{21}^A \mu_{12}^B &  \mu_{22}^A \mu_{11}^B  & \mu_{22}^A \mu_{12}^B \\
	\mu_{21}^A \mu_{21}^B  &  \mu_{21}^A \mu_{22}^B  & \mu_{22}^A \mu_{21}^B & \mu_{22}^A \mu_{22}^B
	\end{array}
	\right]}^{
	\begin{array}{@{}c@{}}
	 \\ \mathstrut
	\end{array}
}.
\]
Now note that condition (\ref{5}) for the gradient system in more convenient notions is
\begin{subequations}\label{5'}
	\begin{eqnarray}
	&& \nu_1=\nu_2 \implies \epsilon_{31}+\epsilon_{32}=\epsilon_{41}+\epsilon_{42},\\  
	&& \nu_3=\nu_4 \implies \epsilon_{13}+\epsilon_{14}=\epsilon_{23}+\epsilon_{24},\\
	 && \nu_5=\nu_6 \implies \epsilon_{21}+\epsilon_{23}=\epsilon_{41}+\epsilon_{43},\\ 
	 && \nu_7=\nu_8 \implies \epsilon_{12}+\epsilon_{14}=\epsilon_{32}+\epsilon_{34}.
	\end{eqnarray}
\end{subequations}
This is trivially satisfied by the aforementioned mutation matrix for independent mutation because of the fact that $\sum_{j=1}^2\mu^A_{ij}=\sum_{j=1}^2\mu^B_{ij}=1$---the normalization condition of probability. In conclusion, \emph{in the presence of independent point mutations at the two different loci, the 2L2A system always behaves as a gradient system in the weak selection limit} and no complicated dynamics like oscillatory or chaotic dynamics can appear. It may be noted, however, that converse of this {\color{black}result need not} be true: Mathematically speaking, mutations that are not independent (mentioned in Sec.~\ref{sec:intro}) may or may not satisfy the gradient condition.

\section{2L2A Non-gradient system}\label{section_4}
Now let us see some of the example where non-convergent dynamical outcomes are possible in 2L2A model in the presence of mutation in the weak selection limit. As is clear from the preceding discussions, we have to consider non-independent mutations as they can result in non-gradient dynamics, and hence oscillatory outcomes. 

Let us consider a simple fitness matrix well studied in literature~{\cite{lewontin1960evolution,hastings1981PNAS}}: 
\begin{equation}\label{57}
\sf{M}=\begin{bmatrix}
k & L_1 & k\\ 
L_2 & K & L_2\\
k& L_1 & k
\end{bmatrix}.
\end{equation} 
It basically says that a genotype homozygous at both locus has the fitness $k$, a genotype homozygous at $A$ ($B$) locus and heterozygous at $B$ ($A$) locus has the fitness $L_1$ ($L_2$), and a genotype heterozygous at both locus has the fitness $K$. Our aim is to illustrate the non-fixed point type dynamics that can appear in such systems in the presence of mutation.

To this end, let us take a mutation matrix that does not satisfies condition (\ref{5}), and hence the corresponding dynamics given by  Eqs.~(\ref{2}) is not a gradient system. Specifically, let us take $\epsilon_{31}=\epsilon_{24}=\epsilon_1$, $\epsilon_{12}=\epsilon_{43}=\epsilon_2$, and other mutation probabilities as zero. Furthermore, in order to achieve analytical tractability, let us choose $\nu_1+\nu_2-\nu_3-\nu_4=0$ and $\nu_5+\nu_6-\nu_7-\nu_8=0$ (note that our choice of ${\sf Q}$ satisfies these) so that the internal fixed point $(p^*,q^*)$ is conveniently located at $(1/2,1/2)$. 

While performing linear stability analysis  of Eqs.~(\ref{2}) about fixed point $(1/2, 1/2)$, the Jacobian comes out to be
\begin{equation}
J=\begin{bmatrix}
\alpha & \beta\\ 
\gamma & \delta\\
\end{bmatrix},
\end{equation}
where $\alpha={(k-K)}/{4}-(\nu_1+\nu_2+\nu_3+\nu_4)/{2}, \beta=(\nu_1-\nu_2-\nu_3+\nu_4)/2, \gamma=(\nu_5-\nu_6-\nu_7+\nu_8)/2$, and $\delta={(k-K)}/{4}-(\nu_5+\nu_6+\nu_7+\nu_8)/2$; and the corresponding eigenvalues are
\begin{equation}\label{22}
\lambda=\frac{(\alpha+\delta)\pm \sqrt{(\alpha-\delta)^2+4\beta\gamma}}{2}\,.
\end{equation}
For the Hopf bifurcation to occur one must have $4\beta\gamma<-(\alpha-\delta)^2$ and it happens at $\alpha+\delta=0$ which can be recast to read $ (k-K)=\sum_{i=1}^{8} \nu_i$. For illustrative purpose, if we now fix $\epsilon_2=0.2$ and  $(k-K)=1$, and work with $\epsilon_1$ as a variable parameter then at $\epsilon_1=0.3$, the Hopf bifurcation should occurs (see Fig.~\ref{fig:1}). 

We observe that in the setting we are working, the condition of the Hopf bifurcation is independent of any constraints on $L_1$ and $L_2$. This has an interesting implication that we now point out: The case $K>L_1, L_2$ and $k>L_1,L_2$ corresponds to strong epistasis in literature~{\cite{hastings1981PNAS}} where it was shown that for oscillatory behaviour to appear in 2L2A model (without mutation), strong epistasis is a necessary condition. However, when non-independent mutation is in action, the limit cycle can appear even in the absence of the strong epistasis because the corresponding Hopf bifurcation condition is dependent only on the values of $k$ and $K$.
{\begin{figure}
		\centering
		\includegraphics[width=85mm, height=70mm]{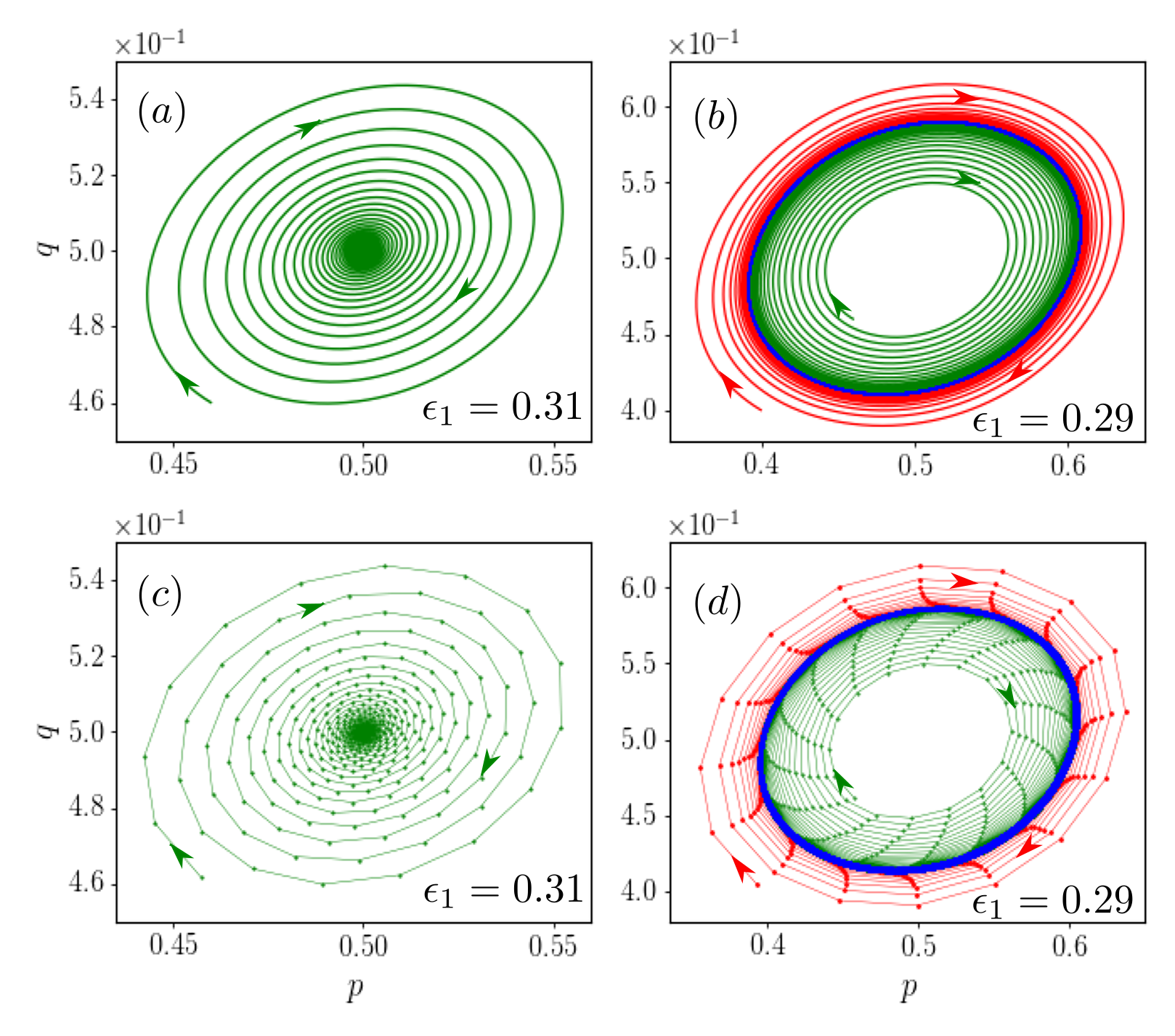}\\
		\caption{(Colour online) {\bf  Mutation drives oscillatory outcomes in non-gradient 2L2A system:} As discussed in Sec.~\ref{section_4}, for illustrative purpose, we take $\nu_1=\nu_4=0.3, \nu_6=\nu_7=0.2$, $\nu_2=\nu_3=\nu_5=\nu_8=0.0$, and $k-K=1.0$. The Hopf bifurcation and the Neimark--Sacker bifurcation (in the time discrete version) occur at $\epsilon=0.3$. Subplots (a) and (b) depict the appearance of limit cycle (blue closed curve) via Hopf bifurcation as $\epsilon_1$ decreases. whereas subplots (c) and (d) exhibit birth of closed invariant cycle (blue closed curve) via Neimark--Sacker bifurcation (we take $s=0.01$). The red and the green trajectories depict evolution of two separate initial conditions. (In the lower panel, the red and the green curves are aid for eyes to follow the discrete orbit.)}
		\label{fig:1}
\end{figure}}

In this context, we recall that the  2L2A gradient system closely approximates the corresponding discrete-time dynamics of  2L2A model in the weak selection limit: Each initial condition approaches a stable fixed point. Somewhat analogous effect is seen corresponding to the limit cycles in discrete-time 2L2A non-gradient system: For the same parameter values and sufficiently small $s$, an invariant closed curve is generated. However, in general, the orbits on the invariant curve need not necessarily be periodic with finite number of periodic points as another possibility is that the orbit on the curve can be everywhere dense (may be due to quasiperiodcity or chaos). As long as $s$ is sufficiently small (we can ignore $\mathcal{O}(s^2)$ and higher order terms), it is easily seen that  the time-discrete equation (which is effectively the Euler forward discretization of the continuous-time equation) under the same parameter values chosen above (while discussing the Hopf bifurcation), undergoes the Neimark--Sacker bifurcation. We do not present the calculations for the sake of avoiding trivial repetition. A numerical evidence of this is showcased in Fig.~\ref{fig:1}.
%


\section{Conclusion}
In conclusion, we have mathematically investigated the two-locus two-allele selection-recombination-mutation dynamics through the replicator equations in the limit of weak selection and weak mutation. We note that when the selection and mutation are sufficiently weak, one can ignore linkage disequilibrium completely and the system becomes two dimensional continuous time differential equation which faithfully approximates the evolutionary dynamics when the dynamics is gradient-like. Our search for the cases of mutation for which the dynamics is gradient-like has led to the result that whenever the point mutation rates at one locus is not affected by what happens at the other, the dynamics is always gradient-like. When the dynamics is not gradient-like, we found that stable limit cycle or stable closed invariant curve can appear. The dynamics on the invariant curve can, in principle, sustain non-convergent solutions like periodic, quasiperiodic, and chaotic orbits. We have also discussed how the dynamical manifestation of epistasis can completely changed in presence of mutation. 

We are tempted to claim that our results imply that, within the paradigm of replicator dynamics of evolutionary process in the weak selection and mutation limit, an observation of oscillatory gametic or allelic frequencies is indirect indication of the existence of mutations that are not independent. An immediate future work that builds on the present paper is to find how the aforementioned results manifest themselves in a finite population where stochastic effects can no longer be ignored. Furthermore, many characteristics like our body size~\cite{Reed2003genome}, height~\cite{Lettre_2008naturegenetics}, and the shape of different organs like the ear~\cite{Adhikari2015natcommun} are controlled by more than two loci of a chromosome. So understanding the evolutionary dynamical outcome of multi-locus systems~\cite{Zhivotovsky1998genetics,Wray2010genome}, in the presence of mutation, is still another natural direction to investigate. While our main result---occurrence of gradient system in the presence of independent mutation---is valid for multi-locus two-allele systems~(see Appendix~\ref{sec:mlma}), the condition for general multi-locus multi-allele systems' becoming gradient-like remains an open problem.

\acknowledgements
The authors thank Yash Varshney for discussions during the initial stages of this work. Sagar Chakraborty acknowledges the support from SERB (DST, govt. of India) through project no. MTR/2021/000119.

\appendix
\section{Multi-locus two-allele gradient system}
\label{sec:mlma}

Let there be a total $n+1$ loci and at $l$th ($l=0,1,2,\cdots,n$) locus two possible alleles be denoted by $A_0^{(l)}$ and $A_1^{(l)}$. The gametes---each represented by a sequence, $A_{\varsigma_0}^{(0)}A_{\varsigma_1}^{(1)}A_{\varsigma_2}^{(2)}A_{\varsigma_3}^{(3)}\cdots A_{\varsigma_j}^{(j)}\cdots A_{\varsigma_n}^{(n)}$, where $\varsigma_j\in\{0,1\}$ $\forall j$---are mathematically sorted in a fashion such that $i$th gamete ($G_i$) is listed at position $i=(\varsigma_0 \times 2^0)+(\varsigma_1 \times 2^1)+(\varsigma_2 \times 2^2)+\cdots+(\varsigma_n \times 2^n)$, thereby helping us to directly associate a gamete number with a gamete. Let the frequency of $G_i$ be $x_i$.

Furthermore, let the fitness of the $G_iG_j$ type of genotype be $w_{ij}$. Assuming that the fitness is independent of which gamete is contributed by which parent, the equality, $w_{ij}=w_{ji}$, follows. The fitness of the $i$th gamete and average fitness of the population are, thus, given by $w_i\equiv\sum_jw_{ij}x_j$ and $\bar w\equiv\sum_j w_jx_j$, respectively.

Now for multi-locus system the selection-recombination evolutionary equation~\cite{nagylaki1999JMB,nagylaki1993Genetics} for the gametes' frequency in the presence of multiplicative mutation which occur during gametic stage \cite{Ewing1981genetics}, takes the form
\begin{equation}
\label{35}
x_i'=\frac{1}{\bar{w}}\sum_{j=0}^{2^n-1}(x_jw_j-D_j)Q_{ji},
\end{equation}
where $Q_{ji}$ corresponds to the probability of mutation from the $j$th gamete to the $i$th gamete. The symbol $D_i$ is understood as follows:

Let us decompose the set of loci $l=\{0,1,2,\cdots,n\}$ into two disjoint sets, $I$ and $J$. Let the probability of recombination with the loci in $I$ inherited from one parent with the loci in $J$ inherited from the other parent be $r_I$.  Now $D_i\equiv\sum_{j}\sum_I r_I(w_{ij}x_i x_j-w_{i_I j_J, j_I i_J}x_{i_Ij_J}x_{j_I i_J})$ represents the measure of linkage disequilibrium for $i$th gamete. $x_{i_Ij_J}$ denotes the frequency of gametes consisting of the genes $i_I$ located in $I$ and the genes $j_J$ located in $J$ inherited from two different parents, where $i_I$ (or $j_J$) corresponds to a vector with component $i_l$ (or $j_l$) for every $l \in I$ (or $l \in J$). In passing, note that, as it should $D_i$  ultimately boils down to $\theta_i w_{14}r D$ for two-locus two-allele scenario, when $w_{14}=w_{23}$, as used in the main text. In consistent notation, $w_{i_I j_J, j_I i_J}$  is the fitness of the genotype formed by the two gametes---one whose frequency is $x_{i_Ij_J}$ and another one whose frequency is $x_{j_Ii_J}$.

Here again we are interested in evolutionary outcome under weak selection and weak mutation. So let us take the form of the fitness and mutation as we have taken for two-locus case (\ref{eqn:selection}).
Under weak mutation, \`a la Nagylaki \cite{nagylaki1993Genetics}, we can show that under weak selection linkage disequilibrium in Eq.~(\ref{35}) goes to quasi-linkage equilibrium within a few generations ($t>t_1$). So we can write Eq.~(\ref{35}) after $t>t_1$ as
\begin{equation}
x'_i=\frac{1}{\bar{w}}\sum_{j=0}^{2^{(n+1)}-1}Q_{ji}x_jw_j.
\end{equation}
Rescaling time $t(= 0, 1, 2...)$ of
generations as $\tau = st$ and taking the limit $s \to 0$, we ultimately come-up with the following differential equation {\cite{hofbauer1998book}}:
\begin{equation}\label{40}
\dot{x}_i=x_i\left[m_i-\bar{m}\right] +\sum_{j=0}^{2^{(n+1)}-1}(\epsilon_{ji}x_j-\epsilon_{ij}x_i).
\end{equation}
We now want to require this equation in terms of the frequency of the alleles.

We denote the frequency of the allele $A_0^{(l)}$, i.e., frequency of the $0$th allele at $l$th locus by $z_l$ and the frequency of the $1$st allele at the $l$th locus by $z_{n+l}$. Since each locus contains two alleles, it is obvious that $z_{n+l}=1-z_l$. At the linkage equilibrium manifold, the gamete frequency is given by just the product of the corresponding allele frequencies~\cite{nagylaki1999JMB}, 
\begin{equation}x_i=\prod_{l=0}^{n}z_{l+\varsigma_ln}\end{equation} where $i=\Sigma_{l=0}^n\varsigma_l2^l$,
 in our notations. $z_{l}$ can be easily calculated by taking the sum over all possible gametes in which the $l$th locus contains the $0$th allele.  Again, by virtue of the numbering convention of the gametes introduced in the beginning of this section, one can write the allele frequency in terms of gamete frequency explicitly as follows:
\begin{equation}\label{34}
z_l=x_0+x_1+\cdots+x_{2^l-1}+x_{2.{(2^l)}}+\cdots+x_{3.{(2^l)}-1}+x_{4.{(2^l)}}+\cdots.\end{equation}

Taking the time derivative of Eq.~(\ref{34}) and using Eq.~(\ref{40}), we ultimately get the evolution equation for allele frequency:
\begin{widetext} 
	\begin{equation}
	\label{12}
	\hspace*{-10mm}	\dot{z}_{l}\equiv \tilde{f}_{l}(\boldsymbol{z})\equiv z_{l}(1-z_{l})\left[\frac{1}{2}\frac{\partial \bar{m}}{\partial z_{l}}  + \nu_{{l}_{0}}\frac{x_0}{z_{l}(1-z_{l})}+ \nu_{{l}_1}\frac{x_1}{z_{l}(1-z_{l})}+...+\nu_{{l}_{2^n}}\frac{x_{2^n}}{z_{l}(1-z_{l})}\right]\equiv z_{l}(1-z_{l})f_{l},
	\end{equation}
where, $\bar{m}=1+s \bar{w}$,\\ and \hspace*{3mm} $\nu_{l_0}\equiv-\left[\epsilon_{0,(2^l)}+\cdots+\epsilon_{0,2(2^l)-1}+\epsilon_{0,3(2^l)}+\cdots+\epsilon_{0,4(2^l)-1}+\cdots\right]$,\\ \hspace*{10mm} $\nu_{l_1}\equiv-\left[\epsilon_{1,(2^l)}+\cdots+\epsilon_{1,2(2^l)-1}+\epsilon_{1,3(2^l)}+\cdots+\epsilon_{1,4(2^l)-1}+\cdots\right]$,\\\hspace*{11.0mm}$\cdots\equiv \cdots$\\\hspace*{5.0mm}$\nu_{l_{(2^l-1)}}\equiv-\left[\epsilon_{{(2^l-1)},(2^l)}+\cdots+\epsilon_{{(2^l-1)},2(2^l)-1}+\epsilon_{{(2^l-1)},3(2^l)}+\cdots+\epsilon_{{(2^l-1)},4(2^l-1)}+\cdots\right],$\\\hspace*{6.0mm} $\nu_{l_{(2^l)}}\equiv+\left[\epsilon_{(2^l),0}+\cdots+\epsilon_{(2^l),(2^l)-1}+\epsilon_{(2^l),2(2^l)}+\cdots+\epsilon_{(2^l),3(2^l)-1}+\cdots\right]$,\\ \hspace*{11.0mm}$\cdots \equiv \cdots$
\end{widetext}
Here we have put a comma in the subscript of $\epsilon_{ij}$ merely for the visual clarity in reading the subscripts.

We know the Shahshahani metric in the gametic space (a simplex of dimension $2^{n+1}-1$) is $g_{ij}(x)={\delta_{ij}}/{x_{i}}$. {\color{black}Now we are interested to calculate the corresponding metric ($\bar{g}_{ij}$, say) in the allelic space, $(z_{1}, z_{2},\cdots,z_{2n})$. Transformation of the coordinates yields:
\begin{equation}\label{37}
\bar{g}_{ij}(z)=\frac{\partial x^{\rho}}{\partial z_{i}}\frac{\partial x^{\sigma}}{\partial z_{j}}g_{\rho \sigma}(x).
\end{equation}}
Evidently, for $i\ne j$, the metric elements $\bar{g}_{ij}(z)$ is zero and for $i=j$ the metric is non-zero such that the metric can eventually be written as
\begin{subequations}
	\begin{eqnarray}
	\bar{g}_{ij}(z)=\frac{\delta_{ij}}{z_{i}(1-z_{i})}.
	\end{eqnarray}
\end{subequations}
Recall that the allelic phase space $\mathbb{P}$ is $(n+1)$-dimensional space, ${\Sigma}^2\times\Sigma^2\times \cdots\times \Sigma^2$, embedded in $\mathbb{R}^{2(n+1)}$; here, $\Sigma^2$ is one dimensional simplex.

We need to consider the interior of $\mathbb{P}$, ${\rm int}\mathbb{P}$ (say); at any point ${\boldsymbol{z}}\in {\rm int}\mathbb{P}$, let the tangent space be denoted by $\mathbb{T}_{\boldsymbol{z}}\mathbb{P}$.  Now for ${\boldsymbol{z}}\in {\rm int}\mathbb{P}$ and $\boldsymbol{\eta}\in \mathbb{T}_{\boldsymbol{z}}\mathbb{P}$, the inner product 
\begin{eqnarray}
\langle\dot{\boldsymbol{z}}, \boldsymbol{\eta} \rangle_{\boldsymbol{z}}=\langle\boldsymbol{\tilde{f}}(\boldsymbol{z}), \boldsymbol{\eta} \rangle_{\boldsymbol{z}}=\bar{g}_{ij}(\boldsymbol{z})\tilde{f}_i(\boldsymbol{z}) \eta_j.
\end{eqnarray}
Considering the metric $\bar{g}_{ij}$, it is obvious that
 \begin{eqnarray}\label{eq:dotz,}
 \langle\dot{\boldsymbol{z}}, \boldsymbol{\eta} \rangle_{\boldsymbol{z}}=\frac{\partial U}{\partial z_i}\eta_i,
 \end{eqnarray}
 if there exists a continuous and differentiable scalar functions $U({\boldsymbol{z}})$ and $\phi(\boldsymbol{z})$ such that $f_i(\boldsymbol{z})={\partial U({\boldsymbol{z}})}/{\partial z_i}+\phi(\boldsymbol{z})$. In other words, this form of $f_i$ is the condition for $\dot{\boldsymbol{z}}=\tilde{\boldsymbol f}$ to be a gradient system.
 
Now, consider the bilinear form $\boldsymbol{H_z \tilde{f}}$:
\begin{equation}\label{42}
\boldsymbol{H_z \tilde{f}}{(\boldsymbol{\xi}, \boldsymbol{\eta})}\equiv \sum_{i,j=1}^{2n}\frac{1}{z_{i}(1-z_{i})}\frac{\partial \tilde{f}_{i}}{\partial z_{j}}\xi_{i} \eta_{j},
\end{equation}
where $\boldsymbol{\xi, \eta} \in \mathbb{T}_{\boldsymbol{z}} \mathbb{P}$. We find that 
\begin{equation}\label{41}
\frac{\partial \tilde{f}_{i}}{\partial z_{j}}=\delta_{ij}(1-2z_{i})f_{i}+z_{i}(1-z_{i})\frac{\partial f_{i}}{\partial z_{j}}.
\end{equation}
Putting this relation in Eq.~ (\ref{42}) and imposing $f_i(\boldsymbol{z})={\partial U({\boldsymbol{z}})}/{\partial z_i}+\phi(\boldsymbol{z})$, we find that $\boldsymbol{H_z \tilde{f}}{(\boldsymbol{\xi}, \boldsymbol{\eta})}$ is symmetric which, in turn, implies
\begin{equation}
\sum_{i, j=1}^{2n}\left(\frac{\partial f_{i}}{\partial z_{j}}-\frac{\partial f_{j}}{\partial z_{i}}\right) \xi_{i}\eta_{j}=0
\end{equation}
for all possible tangent vectors $\boldsymbol{\xi, \eta}\in \mathbb{T}_{\boldsymbol{z}}\mathbb{P}$. With $\boldsymbol{\xi}$=$\mathbf{e_i-e_k}$ and $\boldsymbol{\eta}$=$\mathbf{e_j-e_k}$ (${\bf e}$'s are the unit basis vectors of $\mathbb{R}^{2(n+1)}$), we obtain
\begin{equation}\label{43}
\frac{\partial f_{i}}{\partial z_{j}}+\frac{\partial f_{j}}{\partial z_{k}}+\frac{\partial f_{k}}{\partial z_{i}}=\frac{\partial f_{i}}{\partial z_{k}}+\frac{\partial f_{k}}{\partial z_{j}}+\frac{\partial f_{j}}{\partial z_{i}}.
\end{equation}
As an aside, one notes that this general condition for 2L2A reduces to the simpler condition, $\partial f_0/\partial z_1=\partial f_1/\partial z_0$ as observed in the main text making the presence of $\phi(\boldsymbol{z})$ redundant in the condition.

We now show that just like the fact that independent point mutation renders the 2L2A model a gradient system, multi-locus two-allele model in presence of independent point mutation is also a  gradient system. To this end, it suffices to ignore $\phi(\boldsymbol{z})$, meaning we have as a special case of Eq.~(\ref{43}): 
\begin{equation}\label{32}
\frac{\partial f_{i}}{\partial z_{j}}=\frac{\partial f_{j}}{\partial z_{i}},\quad\forall j\in\{0,1,2,\cdots,n\},
\end{equation} 
at all values of ${\boldsymbol{z}}$. Thus, we explicitly have
\begin{widetext}  
\begin{subequations}\label{31}
	\begin{eqnarray}
	&&\hspace*{-15mm}\frac{\partial f_i}{\partial z_j}=\left[(\nu_{i_{2^j}}-\nu_{i_0})(x_0+x_{2^j})+(\nu_{i_{2^j+1}}-\nu_{i_1})(x_0+x_{2^j+1})+\cdots+(\nu_{i_{2^i}}-\nu_{i_{2^j+2^i}})(x_{2^i}+x_{2^j+2^i})+\cdots\right]\frac{1}{z_i (1-z_i)}\nonumber\\&&+\frac{1}{2}
	\frac{\partial^2 \bar{m}}{\partial z_j \partial z_i},\\&& \hspace*{-15mm}
	\frac{\partial f_j}{\partial z_i}=\left[(\nu_{j_{2^i}}-\nu_{j_0})(x_0+x_{2^i})+(\nu_{j_{2^i+1}}-\nu_{j_1})(x_0+x_{2^i+1})+\cdots+(\nu_{j_{2^j}}-\nu_{j_{2^i+2^j}})(x_{2^j}+x_{2^i+2^j})+\cdots\right]\frac{1}{z_j (1-z_j)}\nonumber\\&&+\frac{1}{2}
	\frac{\partial^2 \bar{m}}{\partial z_i \partial z_j}.  
	\end{eqnarray}
\end{subequations}
\end{widetext}
Owing to the independence of $x_i$'s, from Eq.~(\ref{31}a) and (\ref{31}b), one can easily say that the condition (\ref{32}) is satisfied when
\begin{equation}\label{38}
\nu_{i_{2^j}}=\nu_{i_0},~~ \nu_{i_{2^j+1}}=\nu_{i_1},\cdots,~~
\nu_{i_{2^i}}=\nu_{i_{2^j+2^i}},\cdots
\end{equation}
and
\begin{equation}\label{39}
\nu_{j_{2^i}}=\nu_{j_0},~~
\nu_{j_{2^i+1}} =\nu_{j_1},\cdots,~~
\nu_{j_{2^j}}=\nu_{j_{2^i+2^j}},\cdots.
\end{equation}

Now let the probability of mutation from $A_i^{(l)}$ allele to $A_j^{(l)}$ ($i,j\in\{0,1\}$ and $i\ne j$) allele at locus $l$ be $\mu^{l}_{ij}$. In independent point mutations, the probability of mutation from one gamete to another gamete is just the product of the corresponding allelic mutation probabilities. So we can write down the explicit expressions of $\nu_{i_0}$ and  $\nu_{i_{2^j}}$ ($\forall j<i$):
\begin{widetext}
\begin{subequations}\label{36}
	\begin{eqnarray}
	&&\hspace*{-14mm}\nu_{i_0}=-\left[\left(\mu^0_{00}\mu^1_{00}\cdots\mu^i_{01}\cdots\mu^{n-1}_{00}\mu^n_{00}\right)+\cdots+\left(\mu^0_{01}\mu^1_{01}\cdots\mu^i_{01}\cdots\mu^{n-1}_{00}\mu^n_{00}\right)+\left(\mu^0_{00}\mu^1_{00}\cdots\mu^i_{01}\mu^{i+1}_{01}\cdots\mu^n_{00}\right)+\cdots\right],~~~~~~~~~\\&&\hspace*{-14mm}
	\nu_{i_{2^j}}=-\left[\left(\mu^0_{00}\cdots\mu^j_{10}\cdots\mu^i_{01}\cdots\mu^n_{00}\right)+\cdots+\left(\mu^0_{01}\cdots\mu^j_{11}\cdots\mu^i_{01}\cdots\mu^n_{00}\right)+\left(\mu^0_{00}\cdots\mu^j_{10}\cdots\mu^i_{01}\mu^{i+1}_{01}\cdots\mu^n_{00}\right)+\cdots\right],
	\end{eqnarray}
\end{subequations}
\end{widetext}
and similarly all the other $\nu$'s in terms of the multiplication of mutation probabilities at each locus. One can check from the expressions above that both $\nu_{i_0}$ and $\nu_{i_{2^j}}$ ultimately boil down to $-\mu^i_{01}$. An intuition behind it may be gained by noting that $\nu_{i_0}$ (and $\nu_{i_{2^j}}$) contains terms which are (apart from an overall negative sign) addition of the probabilities of mutation from $0$th (and ${2^j}$th) gamete to the gametes which do not contribute to the allele frequency, $z_i$. Actually, the sum of the latter gametes' frequencies give $1-z_i$. Consequently, we can effectively consider the sum under question as the probability of mutation from $A_0^{(i)}$ allele to $A_1^{(i)}$ allele at the $i$th locus. Likewise, other conditions [(\ref{38}) and (\ref{39})] are also satisfied trivially in the presence of independent mutations. So we can conclude that in the presence of independent point mutations, the multi-locus two-allele system is gradient-like.
\section{Local Lyapunov function}
\label{sec:llf}
Here we show that the function, $V(p,q)\equiv\exp (\bar{m})p^{2\nu_1}(1-p)^{2\nu_3}q^{2\nu_5}(1-q)^{2\nu_7}$, is a local Lyapunov function for the gradient system (\ref{6}). First we note that the value of the function $V(p,q)$ is always positive for all  values of $p$ and $q$ in ${\rm int}\mathbb{P}$. Next we show below that the function has local maxima (minima) at the stable (unstable) fixed points of Eqs.~(\ref{6}).

Considering the alternate form of Eqs.~(\ref{6}), viz., Eqs.~(\ref{eq:7}), we see that every fixed points $(p^*,q^*)\in {\rm int}\mathbb{P}$ corresponds to  ${\partial V}/{\partial p}={\partial V}/{\partial q}=0$. For later use, we write below the explicit expressions of the second derivative of $V(p,q)$:
\begin{subequations}\label{eq:v2}
	\begin{eqnarray}
	&&\frac{\partial^2V}{\partial p^2}\bigg|_{(p^*,q^*)}=\frac{2V\left[m_A-\nu_1-\nu_3\right]}{p^*(1-p^*)},
	\\&& \frac{\partial^2V}{\partial q^2}\bigg|_{(p^*,q^*)}=\frac{2V\left[m_B-\nu_5-\nu_7\right]}{q^*(1-q^*)},\\&&
	\frac{\partial^2V}{\partial p \partial q}\bigg|_{(p^*,q^*)}=V\bar{m};
	\end{eqnarray}
\end{subequations}
where $m_A \equiv \frac{\partial}{\partial p}\left[\frac{p(1-p)}{2}\frac{\partial \bar{m}}{\partial p}\right]\bigg|_{(p^*,q^*)}$  and $m_B \equiv \frac{\partial}{\partial q}\left[\frac{q(1-q)}{2}\frac{\partial \bar{m}}{\partial q}\right]\bigg|_{(p^*,q^*)}$.

Now, in the process of the linear stability analysis, the Jacobian of the linearized system at ($p^*,q^*$) is given by
\begin{equation}\label{1}
J_1\equiv
\begin{bmatrix}
m_A-\nu_1-\nu_3       & 2p^*(1-p^*)\bar{m} \\
2q^*(1-q^*)\bar{m}      & m_B-\nu_5-\nu_7 \\
\end{bmatrix}
\equiv
\begin{bmatrix}
a       & b \\
c      & c\\
\end{bmatrix} 
\end{equation}
and the corresponding eigenvalues are
\begin{equation}
\lambda=\frac{1}{2}\left[(a+d)\pm \sqrt{(a-d)^2+4bc}\right].
\end{equation}
Obviously, a necessary condition for the fixed point $(p^*,q^*)$ to be stable is $(a+d)<0$, which implies
\begin{subequations}\label{24}
	\begin{align}
	\text{either}~~ \nu_1+\nu_3>m_A\\~~\text{or}~~~~~ \nu_5+\nu_7>m_B.
	\end{align}
\end{subequations}
This condition in the light of Eqs.~(\ref{eq:v2}) means that
\begin{subequations}\label{25}
	\begin{eqnarray}
	&& \text{either}~~ \frac{\partial^2V}{\partial p^2}\bigg|_{(p^*,q^*)}<0\\&&~~\text{or}~~~~~ \frac{\partial^2V}{\partial q^2}\bigg|_{(p^*,q^*)}<0.
	\end{eqnarray}
\end{subequations}
Finally, the sufficient condition for the stability of the fixed point is $(a+d)^2>(a-d)^2+4bc$, which boils down to
\begin{eqnarray}\label{26}
\left[m_A-\nu_1-\nu_3\right]\left[m_B-\nu_5-\nu_7\right]>4p^*q^*(1-p^*)(1-q^*)\bar{m}^2,
\nonumber\\
\end{eqnarray}
which in the light of Eqs.~(\ref{eq:v2}) implies that
\begin{eqnarray}\label{eq:v22}
\left[\frac{\partial^2V}{\partial p^2}\bigg|_{(p^*,q^*)}\right]\left[\frac{\partial^2V}{\partial q^2}\bigg|_{(p^*,q^*)}\right] >	\left[\frac{\partial^2V}{\partial p \partial q}\bigg|_{(p^*,q^*)}\right]^2.\quad
\end{eqnarray}
{\color{black}Conditions (\ref{25}) and (\ref{eq:v22})} together imply that the stable fixed point corresponds to local maximum of the function $V(p,q)$. Similar  calculation shows that the unstable fixed point corresponds to local minimum and saddle fixed point corresponds to saddle point of the function $V(p,q)$. Furthermore, we already know from Eq.~(\ref{9}) that $\dot{V}>0$ except at the fixed points where it vanishes. Hence, it qualifies for the local Lyapunov function.

Actually, in the more conventional sense~\cite{jordan2007oxford}, the local Lyapunov function near stable a fixed point, $(p^*,q^*)$, is $V(p^*,q^*)-V(p,q)$ and near unstable a fixed point, $(p^*,q^*)$, is $V(p,q)-V(p^*,q^*)$.
\bibliography{Chakraborty_etal}

\end{document}